\documentclass[%
 reprint,
 amsmath,amssymb,
 aps,
]{revtex4-2}

\usepackage[usenames]{color}
\usepackage{graphicx}
\usepackage{dcolumn}
\usepackage{bm}

\begin{document}

\preprint{APS/123-QED}

\title{Plasmonic dichroism and all-optical magnetization switching in nanophotonic structures with GdFeCo}

\author{Polina E. Zimnyakova}
\affiliation{Moscow Institute of Physics and Technology, National Research University, Dolgoprudny, Moscow, 141701 Russia}
\affiliation{Russian Quantum Center, Moscow, Russia}

\author{Daria O. Ignatyeva}
\affiliation{Faculty of Physics, M.V. Lomonosov Moscow State University, Moscow, Russia
}
\affiliation{V.I. Vernadsky Crimean Federal University, Simferopol, Russia}
\affiliation{Russian Quantum Center, Moscow, Russia}

\author{Andrey N. Kalish}
\affiliation{Faculty of Physics, M.V. Lomonosov Moscow State University, Moscow, Russia
}%
\affiliation{Russian Quantum Center, Moscow, Russia}

\author{Xiufeng Han}
\affiliation{Beijing National Laboratory for Condensed Matter Physics, Institute of Physics, University of Chinese Academy of Sciences, Chinese Academy of Sciences, Beijing, China}

\author{Vladimir I. Belotelov}
\affiliation{Faculty of Physics, M.V. Lomonosov Moscow State University, Moscow, Russia
}
\affiliation{Russian Quantum Center, Moscow, Russia}
\affiliation{V.I. Vernadsky Crimean Federal University, Simferopol, Russia}

\date{\today}

\begin{abstract}
We report on a phenomenon of plasmonic dichroism observed in magnetic materials with the transverse magnetization under the excitation of the surface plasmon polariton waves. The effect originates from the interplay of the two magnetization-dependent contributions to the material absorption, both of which are enhanced under plasmon excitation. Similar to the recently discovered effect of a  all-optical helicity-dependent magnetization switching, this effect provides a possibility to perform a deterministic magnetization switching to the desired state. We show by electromagnetic modeling that laser pulses exciting counter-propagating plasmons can be used to write +M or -M state in a deterministic way independent on the initial magnetization state. The presented approach applies to various ferrimagnetic materials exhibiting the phenomenon of all-optical switching of thermal nature and broadens the horizons of their applications in data storage devices.
\end{abstract}

\maketitle



Ultrafast magnetization switching via femtosecond laser pulses is promising for fast and dense magnetic information recording~\cite{kimel2020fundamentals}. A phenomenon of all-optical magnetization switching was recently demonstrated in several magnetic materials, including GdFeCo films~\cite{vahaplar2009ultrafast,igarashi2020engineering,gschneidner2004handbook,ostler2012ultrafast}, Co/Pt multilayers~\cite{gorchon2017single}, CrI$_3$ monolayers~\cite{kudlis2021all}, Pt/Co/Gd stacks~\cite{lalieu2017deterministic}, cobalt-substituted garnet films ~\cite{stupakiewicz2017ultrafast}. In most of these cases, especially in widely studied GdFeCo films, thermal mechanism plays a decisive role in the process of all-optical magnetization switching~\cite{ostler2012ultrafast,lu2018roles}. This leads to several consequences practically important for the data storage applications. First, the magnetization switching process is nearly insensitive to the laser polarization and wavelength. Second, it is important that all-optical magnetization switching is a threshold effect that can be observed only if the absorbed energy exceeds a certain value~\cite{khorsand2012role}. This means, one may tune a laser power to the near-threshold value so that due to the magnetic circular dichroism phenomenon~\cite{vahaplar2012all,khorsand2012role} right circular polarization would switch +M state and would not affect -M state, and vice versa. This effect is called the all-optical helicity-dependent magnetic switching (AO-HDS)~\cite{stanciu2007all,wang2018all,khorsand2012role}. Due to AO-HDS, one may deterministically write the desired +M or -M state using the laser pulse of the corresponding polarization. Recent studies show that such threshold for magnetization switching also opens new horizons for the magnetization switching in the multilayered structures~\cite{ignatyeva2019plasmonic,borovkova2021layer}.

AO-HDS could be observed only for the circularly polarized light. For the linear light polarization, AO-HDS vanishes. At the same time, linear polarization is used to excite propagating~\cite{ignatyeva2019plasmonic} or localized surface plasmons~\cite{liu2015nanoscale} which provide exciting possibilities for multilayered information writing~\cite{ignatyeva2019plasmonic} or reduction of the bit size down to~50 nm~\cite{liu2015nanoscale}. Moreover, plasmonic structures significantly enhance the efficiency of light interaction with in-plane magnetized films where the phenomenon of  all-optical~\cite{ostler2012ultrafast} magnetization switching was observed recently. 

In the present manuscript, we analyze and describe an analog of AO-HDS that could be observed in plasmonic structures using linearly polarized incident light. 

The interaction of light with magnetic materials depends on magnetization so that the variations of the polarization, intensity, and absorption of light are observed under the application of different magnetic fields. Here we focus on the phenomenon of the magnetization-dependent absorption of light. A well-known special case of this phenomenon is magnetic circular dichroism which reveals itself as a different absorption of right and left circularly polarized light in a magnetic medium. Let us analyze the physical origins of magnetization-dependent absorption in more detail.

The power $p$ absorbed by a unit of volume is given by the following equation:
\begin{equation}\label{JLl} 
p = \frac{\omega}{8\pi}\frac{\varepsilon_{ji}-\varepsilon^{*}_{ij}}{2i}  E_{i}E_{j}^{*}, 
\end{equation}
where $\varepsilon_{ij}$ is the permittivity tensor and $^{*}$ is a conjugation symbol (see Appendix), and $E_i$ are the components of the electric field $\mathbf{E}$ of optical radiation. Thus, the absorption coefficient of the structure $A$ may be obtained as $A=|S_z|^{-1}\int p\, \mathrm{d}V$, where $S_z$ is $z$-component of the Poynting's vector of the incident light, and the $z$-axis is normal to the structure's surface. 

An isotropic magnetic material is described by the permittivity tensor of the form $\varepsilon_{ij}=\varepsilon \delta_{ij} - i\epsilon_{ijk}g_k$, where $\delta_{ij}$ is the Kronecker delta, $\epsilon_{ijk}$ is the Levi-Civita symbol, $\varepsilon$ is the diagonal component of the permittivity tensor, and $g_i$ is the gyration vector that is proportional to the magnetization vector. For a smooth magnetic film the absorption coefficient therefore could be expressed as:
\small
\begin{equation}\label{Absorp} 
    A =  \frac{k_0}{n_{sup}\cos\theta}\bigg[\int \varepsilon^{''}|\mathbf{E}|^2\, \mathrm{d}z + \int i \mathbf{g}^{''}[\mathbf{E}\times\mathbf{E^*}]\, \mathrm{d}z\bigg],
\end{equation} \normalsize
where $\theta$ is the angle of incidence, $n_{sup}$ is the refractive index of the material of the incident medium (superstrate), $k_0$ is the vacuum wave vector of the incident light, $\varepsilon^{''}$ and $g^{''}_i$ are the imaginary parts of the diagonal components of the permittivity tensor and the gyration vector, respectively, and $E_{i}$ are components of the light electric field normalized to the field of the incident wave.

As can be seen from Eq.~\eqref{Absorp}, the absorption of light in a magnetic material has two contributions: $A=A_{\varepsilon"}+A_{g"}$ associated with imaginary parts of permittivity and gyration, correspondingly. Therefore, absorption variation  under the material remagnetization $\delta A=A(+M)-A(-M)$ also has two corresponding inputs, $\delta A_{\varepsilon"}$ and $\delta A_{g"}$. According to Eq.~\eqref{Absorp}, $\delta A_{\varepsilon"}$ contribution appears for the light with arbitrary polarization if the material remagnetization changes the light intensity $|\mathbf{E}|^2$ inside it. The contribution of $\delta A_{g"}$ is non-zero only for the circular polarization of light. The sign of $\delta A_{g"}$ changes to the opposite if the magnetization direction changes. For example, a well-known phenomenon of the magnetic circular dichroism is caused purely by this $\delta A_{g"}$ contribution while $\delta A_{\varepsilon"}$ impact is negligible. We argue that in the plasmonic structures, the situation drastically changes and both contributions become essential.

\begin{figure}
    \centering
    \includegraphics[width=1.0\linewidth]{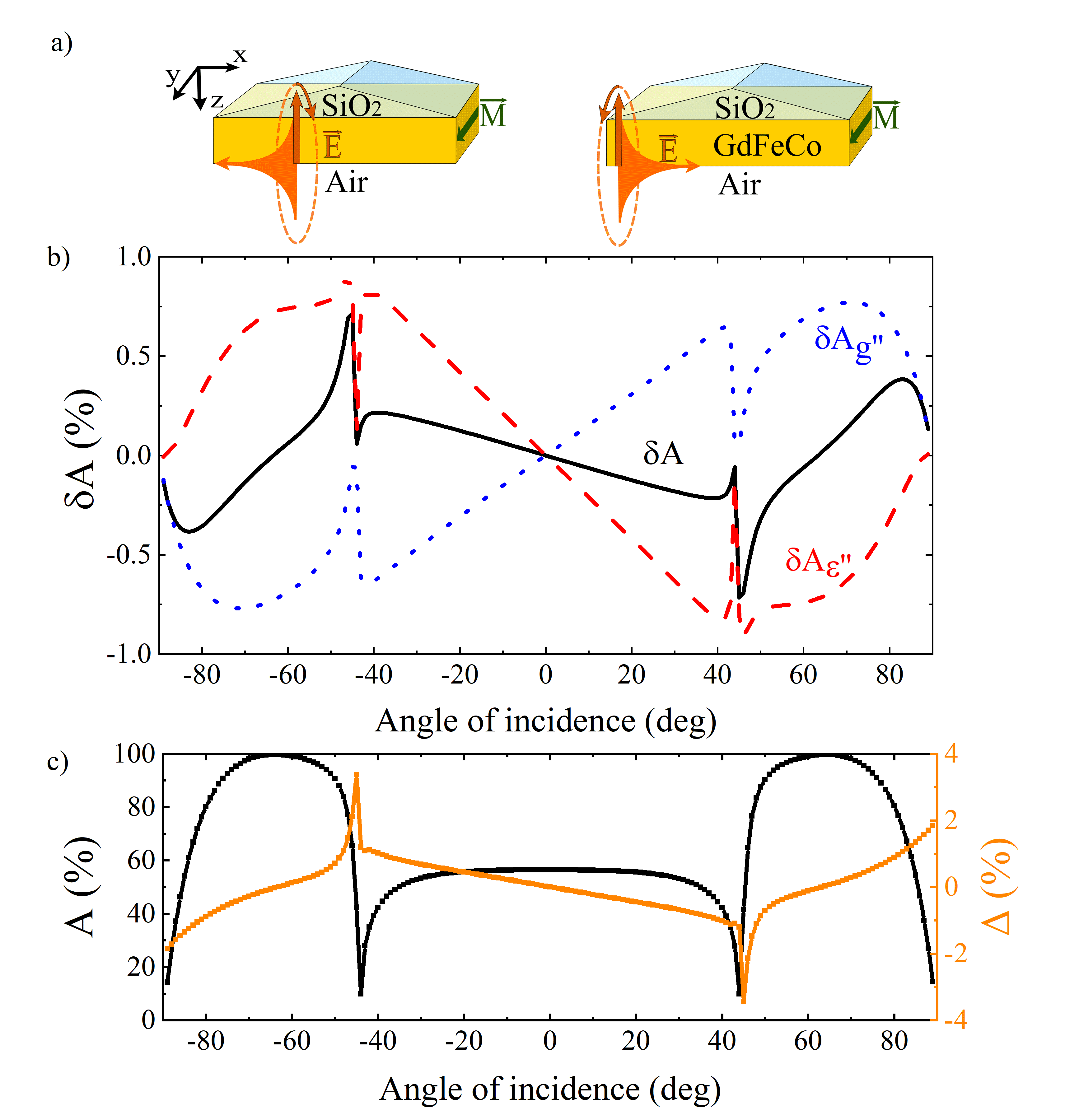}
    \caption{Plasmonic dichroism in a GdFeCo film. (a) Schematic representation of the excited plasmons. (b) Calculated magnetization-induced absorption difference $\delta A$ (black line), and two contributions to the induced absorption: $\delta$A$_{\varepsilon"}$ (red dashed line) and $\delta$A$_{g"}$ (blue dashed line) calculated for the opposite directions of the GdFeCo magnetization. (c) Absorption in the plasmonic GdFeCo structure. Light wavelength is 800 nm.}
    \label{fig:A1A2}
\end{figure}

To illustrate this, we performed numerical simulations of the surface plasmon excitation in a 20-nm-thick GdFeCo film thickness with transverse magnetization via a glass prism. Optical absorption along with both terms in Eq.~\eqref{Absorp} was calculated for different angles of incidence using the rigorous coupled-wave analysis (RCWA~\cite{li2003fourier}) method, see Fig.~\ref{fig:A1A2}. The laser wavelength was chosen 800 nm and GdFeCo permittivity and gyration were taken from~\cite{khorsand2012role}. Fig.~\ref{fig:A1A2}b shows magnetization-dependent variation of absorption $\delta A$ and its contributions $\delta A_{\varepsilon"}$ and $\delta A_{g"}$. One may see that both contributions have similar values, and maximal values of $\delta A _{g"}$ correspond to the plasmon-induced absorption peak in the total internal reflection angular region.

Surface plasmon polarition electric field components at the semi-infinite metal-dielectric interface inside a transversally magnetized metal can be expressed as~\cite{sukhorukov2011terahertz}:
\begin{equation}
    \mathbf{E}_{\mathrm{SPP}\pm x}=\begin{pmatrix}
1 \\
0\\
-i\frac{\pm\beta_g}{\gamma_m}
\end{pmatrix} E_0 \exp(i(\pm\beta_g x - i \gamma_m z - \omega t)),
\label{SPP_efield}
\end{equation}
where $\beta_g$ is a propagation constant, $\pm$ sign indicates propagation in $+x$ or $-x$ direction, $\gamma_m=\sqrt{\beta_g^2-\varepsilon_m k_0^2}$ is localization coefficient, $k_0=\omega/c$ is a vacuum wavevector, $\omega$ is the SPP frequency, $z=0$ denotes the plane corresponding to a dielectric($z>0$)/metal($z<0$) interface. 

\begin{figure}
    \centering
    (a)~~~~~~~~~~~~~~~~~~~~~~(b)~~~~~~~~~~~~~~~~~~~~~~(c)\\
    \includegraphics[width=0.32\linewidth]{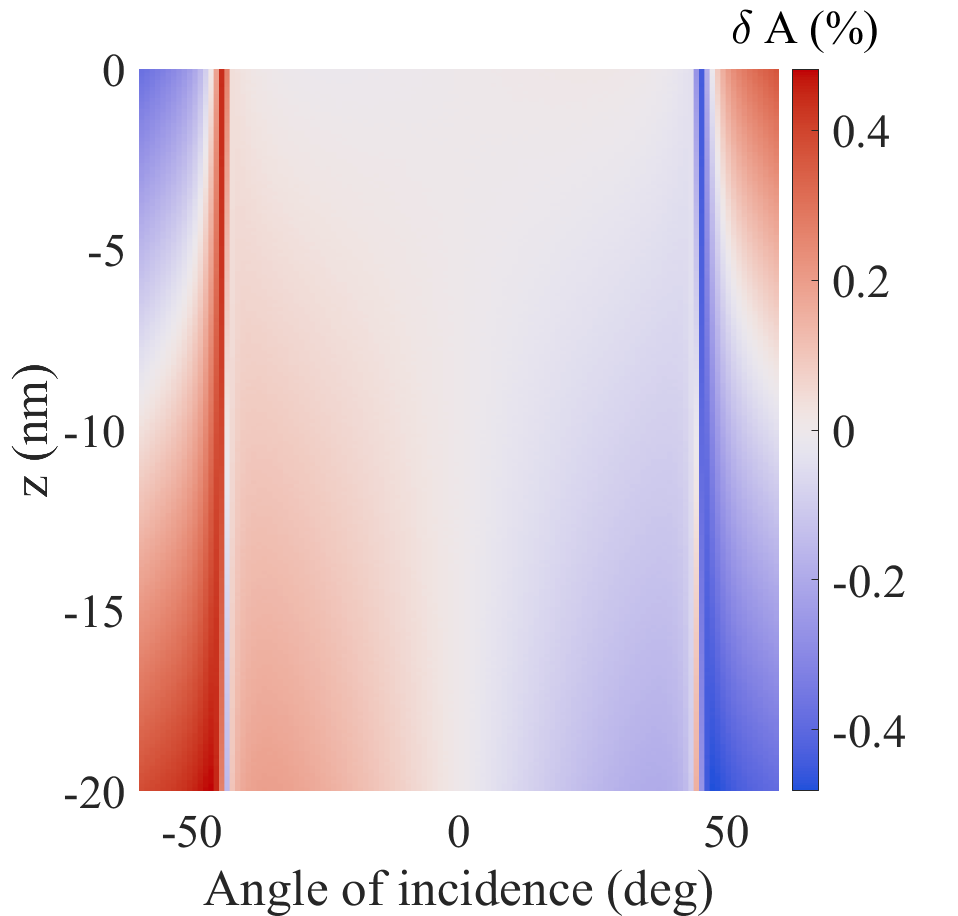} \includegraphics[width=0.32\linewidth]{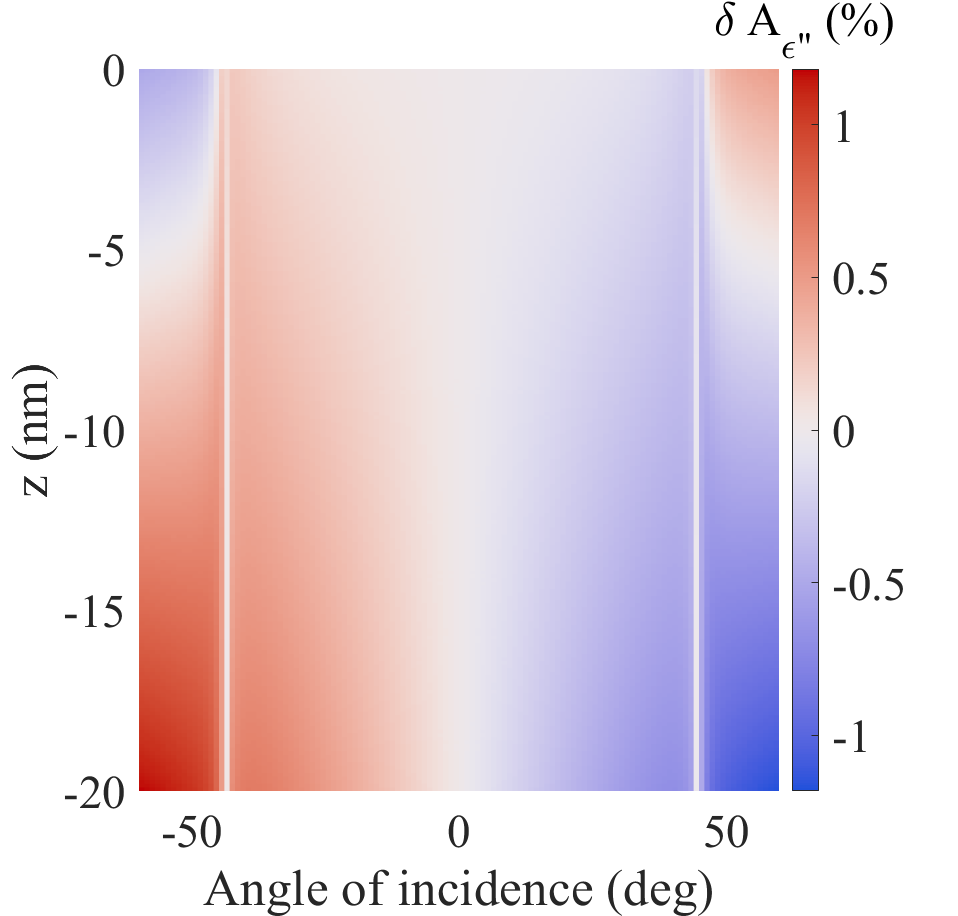} \includegraphics[width=0.32\linewidth]{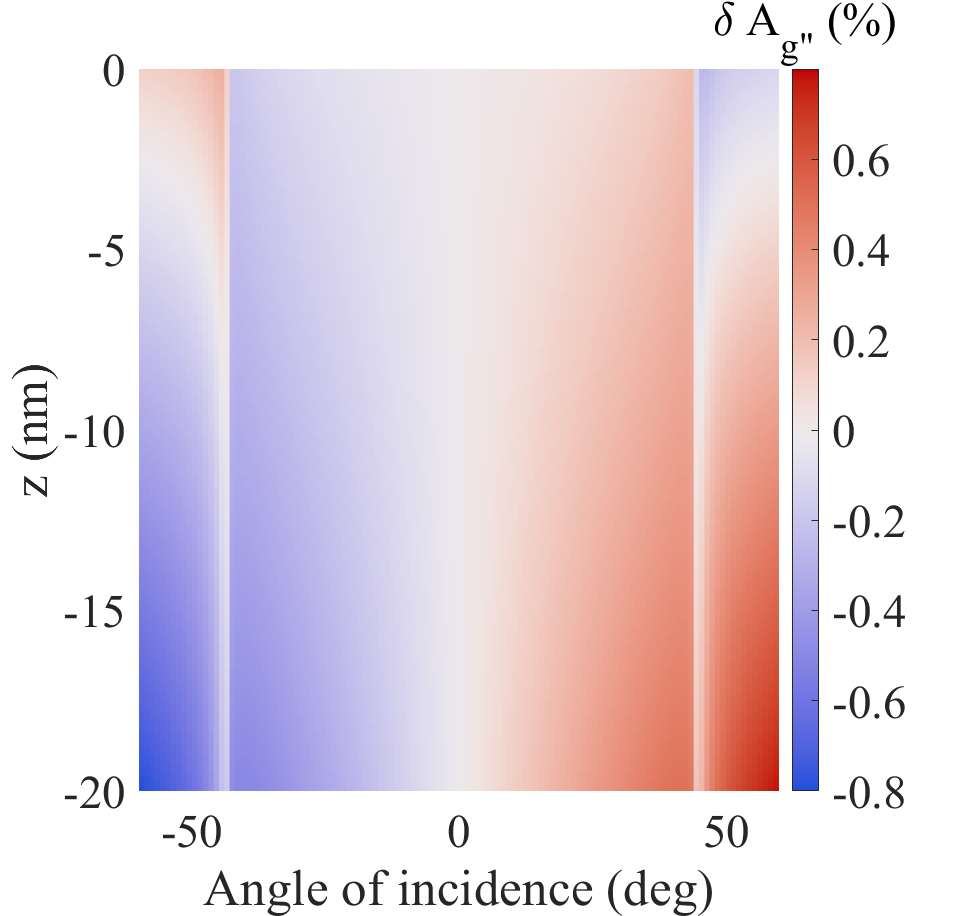}
    \caption{Role of surface plasmons in a plasmonic dichroism. Distributions of (a) $\delta A$  (b) $\delta$A$_{\varepsilon"}$ (c) $\delta$A$_{g"}$ inside GdFeCo film are shown for various angles of incidence and light wavelength of 800~nm.}
    \label{fig:dA1A2}
\end{figure}

It is well-known that the light intensity in the plasmonic structure changes during remagnetization. This is a manifestation of the transverse magneto-optical Kerr effect (TMOKE) in reflected or transmitted light which was observed in various plasmonic and guided-wave supporting structures~\cite{maccaferri2015resonant,torrado2010magneto,khramova2019resonances,voronov2020magneto,bsawmaii2022magnetic}. This effect originates from the non-reciprocal change of the mode propagation constant $\beta_g$ which in the case of the smooth plasmonic films is~\cite{sukhorukov2011terahertz}:
\begin{equation}\label{beta}
    \beta_g = k_0 \sqrt{\frac{\varepsilon_m\varepsilon_d}{\varepsilon_m+\varepsilon_d} }\left(1+g_y \frac{(-\varepsilon_m\varepsilon_d)^{\frac{3}{2}}}{\varepsilon_m^2-\varepsilon_d^2}\right)),
\end{equation}
and in a more general case could be represented as $\beta = \beta_0+g_y\Delta \beta$. This linear in $g$ term shifts the resonance position determined by $\beta_g$ and leads to the changes of the mode excitation efficiency under the fixed light parameters (wavelength and angle of incidence). This gives rise to a contribution of $\delta A_{\varepsilon"}$ (see Fig.~\ref{fig:A1A2}b and Fig.~\ref{fig:dA1A2}b). 

Another important point is that although surface plasmons are excited by the linearly polarized light, the electric field of SPP is elliptically polarized according to Eq.~\eqref{SPP_efield}. Thus, $y$-component of the vector product $[\mathbf{E}\times\mathbf{E^*}]_y$ is nonzero~\cite{belotelov2012inverse,chekhov2018surface} and it has the opposite direction for the two counter-propagating plasmons~\cite{im2019all} (see Fig.~\ref{fig:A1A2}a). This causes the emergence of the magnetization-dependent $\delta A_{g"}$ contribution to the material absorption under surface plasmon excitation (see Fig.~\ref{fig:A1A2}b and Fig.~\ref{fig:dA1A2}c).

 Therefore, in the case of plasmonic modes, these two effects work simultaneously leading to $\delta A$ values comparable to the ones achieved due to the magnetic circular dichroism. Such phenomenon may be treated as a 'plasmonic dichroism'. 
 Therefore, the excited SPP modes provide the opposite sign of the magnetization-dependent absorption component $\delta A$. Fig.~\ref{fig:A1A2}c shows the value of the plasmonic dichroism calculated as:
\begin{equation}
    \Delta = \frac{A(+M) - A(-M)}{\frac{1}{2}(A(+M) + A(-M))},
\end{equation}
similar to magnetic circular dichroism~\cite{khorsand2012role}. One may see that for the fixed wavelength, one may choose the two angles of incidence that provide the opposite values of $\Delta$.  
 
 Non-zero $\Delta$ means that if a plasmon is excited in a magnetic structure, the amount of absorbed energy for domains magnetized in different directions will be different. If a material exhibits the all-optical switching, then, similarly to AO-HDS, there exists a window of laser fluencies in which the laser pulse excites the surface plasmon propagating in a certain direction that switches the material magnetization for the one state and makes no impact on the other, and vice versa. Variation of absorption, $\delta A$ is odd in the angle of incidence. Therefore, to turn the situation upside down and reverse the magnetization back, one may change the light angle of incidence to the opposite one. 

Although the physical origin of the plasmonic dichroism is well illustrated using a smooth film and a prism coupler, various nanophotonic structures exhibit similar properties. Both the magneto-optical intensity effect (Eq.~\eqref{beta}) and the non-zero  product $[\mathbf{E}\times\mathbf{E^*}]$ could be observed for the other types of TM-polarized propagating guided~\cite{krichevsky2021selective} or surface waves. Plasmonic nanogratings could also be used instead of the smooth films with prism couplers.

\begin{figure}
   
    (a)\\
    \includegraphics[width=0.2\linewidth]{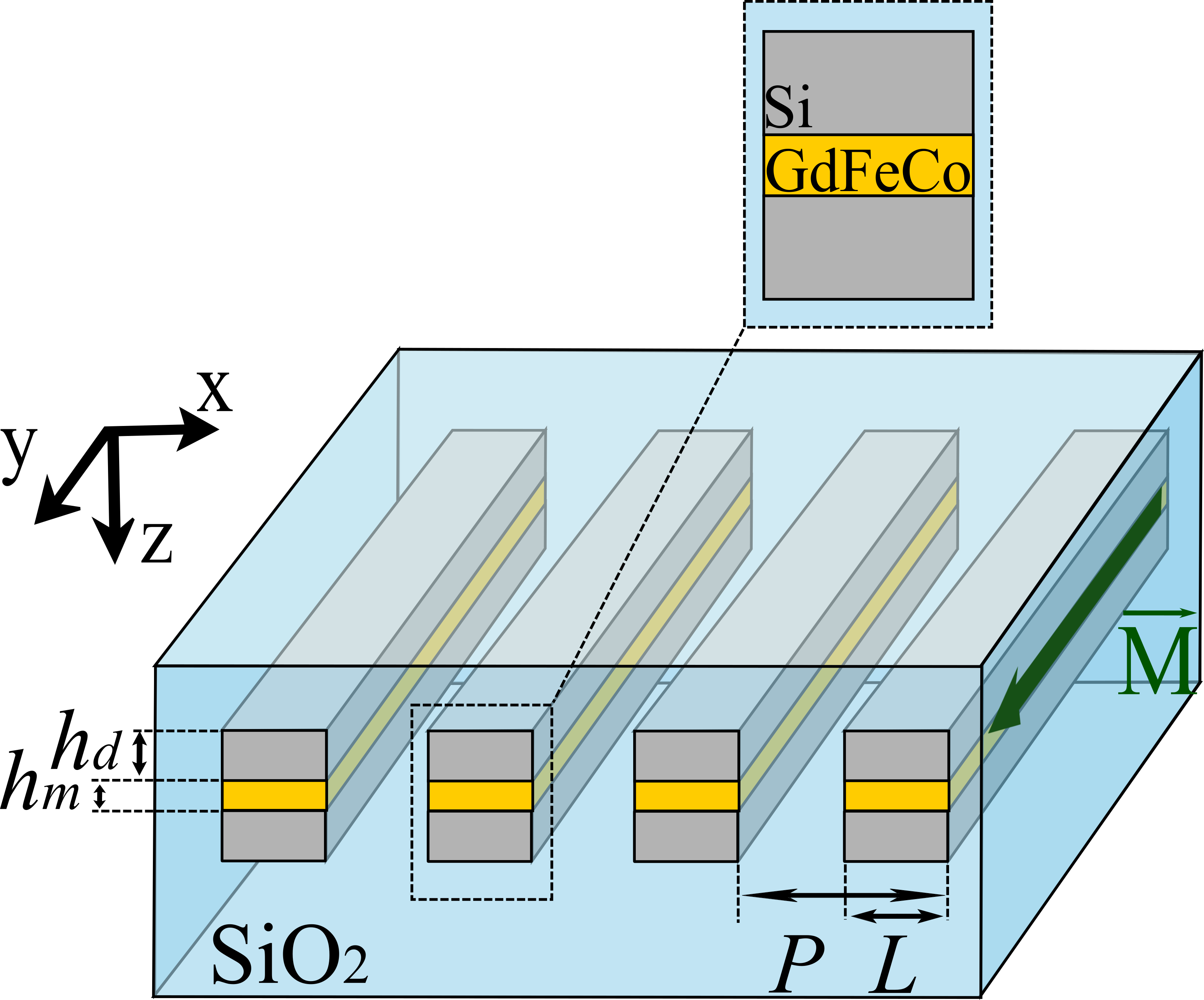}\\
    (b)~~~~~~~~~~~~~~~~~~~~~~~~~~~~~~~~~~~~(c)\\
    \includegraphics[width=0.49\linewidth]{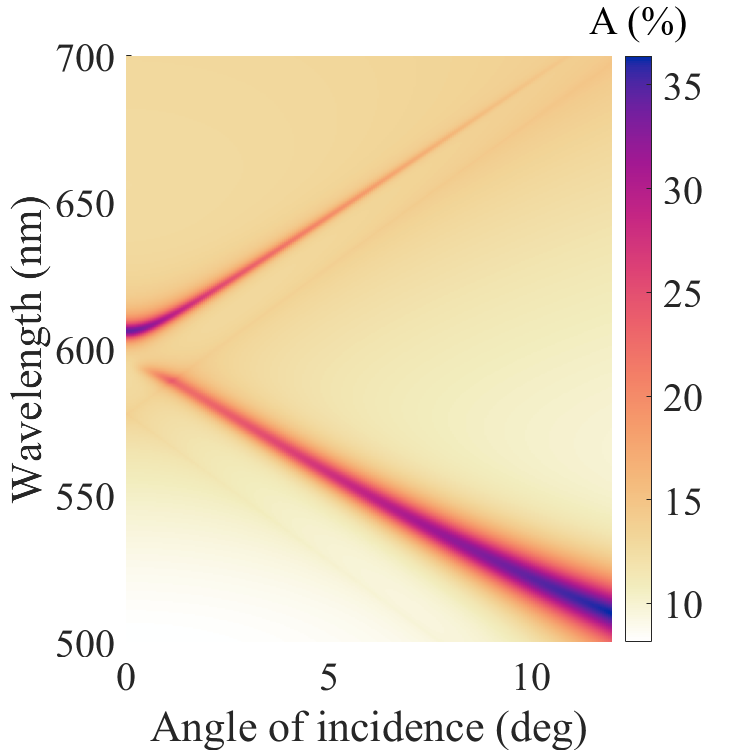}
    \includegraphics[width=0.49\linewidth]{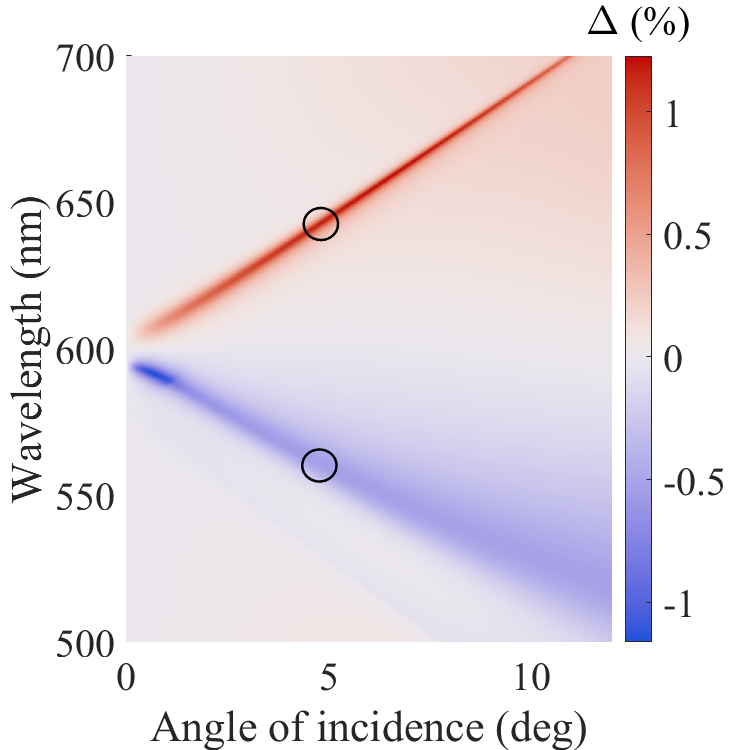}
    \caption{Plasmonic dichroism in a hybrid GdFeCo-Si grating. (a) Schematic representation of the structure. (b-d) Color plots for calculated: (b) SPP electromagnetic field distribution $|H_y|$ at lambda= 606 nm and angle of incidence = 0 deg, (c) absorption (d) plasmonic dichroism $\Delta$ in the considered nanoplasmonic grating.}
    \label{fig:structure and AbsMCD}
    \end{figure}

 To illustrate the latter, we have performed a numerical study of the GdFeCo nanostructure, in which plasmonic modes are excited using a grating.  The following nanostructure parameters are selected: the grating period is $P = 400$~nm, the GdFeCo stripe thickness is $h_{m} = 7$~nm and width $L = 225$~nm, the thickness of Si stripes is $h_{d} = 20$~nm. We consider a one-dimensional grating made of thin GdFeCo stripes to support the long-range surface plasmon modes~\cite{maier2007plasmonics}. The grating is surrounded by Si layers to enhance the efficiency of the SPP mode excitation. The whole structure is embedded in a SiO$_2$ matrix. Schematic representation of this structure is in Fig.~\ref{fig:structure and AbsMCD}a.

Fig.~\ref{fig:structure and AbsMCD}a shows the absorption spectra of nanoplasmonic structure with GdFeCo. One may see two SPP branches which are accompanied by absorption peaks. In contrast to a prism, grating coupler enables excitation of SPPs propagating in $+x$ and $-x$ directions at different wavelengths and fixed angle of incidence in accordance with the phase-matching condition $\pm\mathbf{\beta}_{g}=2\pi/\lambda \sin(\theta)\pm2\pi/P$~\cite{khramova2019resonances}. Consequently, in the considered nanoplasmonic structure a deterministic magnetization switching can be also performed based on the plasmonic dichroism phenomenon. 

The plasmonic dichroism phenomenon makes it possible to use the laser pulses exciting counter-propagating plasmons to write +M or -M state in a deterministic way independent of the initial magnetization state. The presented approach applies to various ferrimagnetic materials exhibiting the phenomenon of all-optical switching of thermal nature. This significantly broadens the horizons of the ferrimagnetic nanostructure applications for data storage devices.

\section*{Funding}
Numerical studies of a plasmonic dichroism were supported by RSF, Grant No. 19-72-10139. Theoretical description of magnetization-dependent absorption was performed in a framework of RFBR Grant No. 18-29-20113. 

\section*{Appendix A}

The Maxwell's equations for the optical waves in the frequency domain have the form:

\begin{equation}\label{Maxwell}\begin{gathered}
    \epsilon_{ijk} \frac{\partial}{\partial x_j} E_k = i k_0 H_i,\\
    \epsilon_{ijk} \frac{\partial}{\partial x_j} H_k = -i k_0 \varepsilon_{il} E_l.
\end{gathered}
\end{equation}

From these equations one can obtain the energy conservation law in the form

\begin{equation}\label{Energy}
    \frac{\partial}{\partial x_i} S_i + p = 0,
\end{equation}
where $S_i=\frac{c}{8\pi} \epsilon_{ijk} \text{Re} (E_j H_k^*)$ is the Poynting vector, and $p$ is given by Eq.~(\ref{JLl}). For the case of the isotropic dielectric tensor Eq.~(\ref{JLl}) takes the form of the conventional Joule-Lenz law:

\begin{equation}\label{JLl0} 
p = \frac{\omega \varepsilon^{''}}{8\pi} |\mathbf{E}|^2. 
\end{equation}


%

\end{document}